# Energy Recovery Linac Based Fully Coherent Light Source

Z.T. Zhao[1, 2*], Z. Wang[1], C. Feng[1, 2], S. Chen[1], L. Cao[1, 3]

[1]Shanghai Advanced Research Institute, Chinese Academy of Sciences, Shanghai 201210, China
[2]University of Chinese Academy of Sciences, Beijing 100049, China
[3]Shanghai Institute of Applied Physics, Chinese Academy of Sciences, Shanghai 201210, China
*e-mail: zhaozhentang@sari.ac.cn

**Abstract**

Energy recovery linac (ERL) holds great promise for generating high repetition-rate and high brightness electron beams. The application of ERL to drive a free-electron laser is currently limited by its low peak current. In this paper, we consider the combination of ERL with the recently proposed angler-dispersion induced microbunching technique to generate fully coherent radiation pulses with high average brightness and tunable pulse length. Start-to-end simulations have been performed based on a low energy ERL (600 MeV) for generating coherent EUV radiation pulses. The results indicate an average brightness over $10^{25}$ phs/s/mm$^2$/mrad$^2$/0.1%BW and average power of about 100 W at 13.5 nm or 20 W with the spectral resolution of about 0.5 meV with the proposed technique. Further extension of the proposed scheme to shorter wavelength based on an ERL complex is also discussed.

**Introduction**

Over the past half-century, remarkable interests and demands of the synchrotron radiation users in the extreme ultraviolet (EUV) and x-ray regime lead to the continuing improvements of the synchrotron radiation facilities (SRs) in four generations, impacting on many disciplines such as physics, chemistry, biology and material science [1]. To date, the 3$^{rd}$ generation light sources, with the radiation mainly generated from the insertion devices, have witnessed an impressive development worldwide, instead of the 1$^{st}$ and the 2$^{nd}$ generation light sources with the radiation mainly emitted from the bending magnets. The average brilliance (B) of the 3$^{rd}$ generation light sources, defined as the photon flux (F) over the transverse photon beam size ($\Sigma_x\Sigma_y$) and divergence ($\Sigma'_x\Sigma'_y$) in 0.1% spectral bandwidth, $B=F/4\pi^2\Sigma_x\Sigma'_x\Sigma_y\Sigma'_y$, is typically $10^{19}$ phs/s/mm$^2$/mrad$^2$/0.1%BW. However, the durations of radiation pulses from storage rings are still too long to measure the atomic motion and structural dynamics on the fundamental time scale of a vibrational period (~100 fs). In addition, the synchrotron radiation pulses are incoherent in both transverse and temporal, limiting the application on many frontier sciences such as high-resolution spectroscopy and imaging experiments.

In order to further improve the brilliance of the synchrotron radiation light source, the diffraction limited storage rings (DLSRs), recognized as one type of the 4$^{th}$ generation light sources, have been developed in the past decades [1-4]. It is known that it helps to provide higher brightness and space coherence with a so-called multi-bend achromat (MBA) storage ring lattice design, that is to decrease the bending angle in each of the dipole bending magnets, allowing stronger focusing by multipole magnets between the bending magnets, instead of the double or triple bend achromat (DBA or TBA) lattice mostly employed in the 3$^{rd}$ generation light sources. The

DLSRs provide high average brightness of about $10^{22}$ phs/s/mm$^2$/mrad$^2$/0.1%BW and high coherent fractions of about 0.1 at photon energy of 10keV, which are enhanced by 3 and 1 orders of magnitude respectively, comparing with those of the 3$^{rd}$ generation light sources.

Free electron lasers (FELs) [5-9], capable of providing high peak power, coherent radiation, are recognized as another revolutionary research tool for various fields. The high-quality electron beam, generated by the linear accelerator, travels through a long undulator line and produces coherent radiation pulses with the peak brightness of $10^{33}$ phs/s/mm$^2$/mrad$^2$/0.1%BW, about 9 order of magnitude higher than that of the 3rd generation light sources. However, the low repetition rate (~100 Hz with copper linac) of FELs leads to an equal average brightness. In order to improve the repetition rate, superconducting linac based free electron lasers, such as FLASH, European-XFEL, LCLS-II and SHINE [10, 11], are built or under-constructing worldwide. With the repetition rate about 3-4 orders of magnitude higher than the normal conducting linac, the superconducting linac based FELs provide higher average brightness radiation pulses and support more users in more potential applications.

Energy recovery linac (ERL) [12-15] is another type of high energy accelerator which in principle combines the advantages of storage ring and linac to profit light sources. The fundamental difference between an ERL and a conventional storage-ring is that the electron beam power can be recycled at near-perfect efficiency to accelerate new electron bunches. The high-quality electron beams with low emittance, low energy spread and especially ultra-short duration (100 fs to several ps) in ERL make it possible to open up new scientific frontiers such as ultrafast dynamics, which is not suitable with storage-ring sources. However, the achieved average current of the ERL (tens mA) [16], mainly limited by the photoinjector and the beam break up criteria in superconducting RF linac (HOM effects) [17, 18], is almost one order of magnitude lower than that of conventional storage-ring light source (about several hundred mA) currently achieved, restricting its application at x-ray regime.

Different types of light sources have their own pros and cons. It is a natural idea to combine the advantages of different techniques to establish a new light source that can provide ultra-short or fully coherent pulses with high repetition rate. For example, several methods have been developed in the last decades to improve the temporal properties of storage rings. Most of these methods employ strong external femtosecond lasers to manipulate the electron beam in the storage ring [19-22]. Requirements for the high peak power and high repetition-rate laser make it challenging to implement these techniques on storage ring in the VUV and x-ray regime. Recently a novel technique termed angler-dispersion induced microbunching (ADM) was proposed for directly imprinting strong coherent microbunching on the electron beam with very small laser-induced energy spread, which significantly reduces the requirements on the laser power [23]. However, the repetition rate is still limited by the damping time of the storage ring, or a complicated scheme called "modulation-antimodulation" beam manipulation technique has to be adopted. Another approach is maintaining a steady state micro bunched beam in the storage ring, namely SSMB [24], which is already under experimental demonstration of the mechanism [25].

Here we report a new method for high repetition rate fully coherent pulse generation by implementing the ADM technique on the ERL. This method takes fully advantages of the ERL with GHz-level high repetition rate and ADM with less requirements on the seed laser power and electron beam current. Simulations with typical parameters of an ERL show that fully coherent pulses with average brightness of about 5-6 orders of magnitude higher than that of the DLSR can be produced, making the proposed method a potential candidate for the next generation light source.

**An EUV light source based on ERL with the ADM scheme**

The layout of the proposed scheme is shown in Fig. 1. A DC gun based photoinjector is used to generate high quality electron beam with low emittance, short pulse durations and high repetition rate. Then the electron beam is transported into the main linac and the downstream recirculating loop for the acceleration and radiation. Finally, the electron beam passes through the main linac to recover the energy.

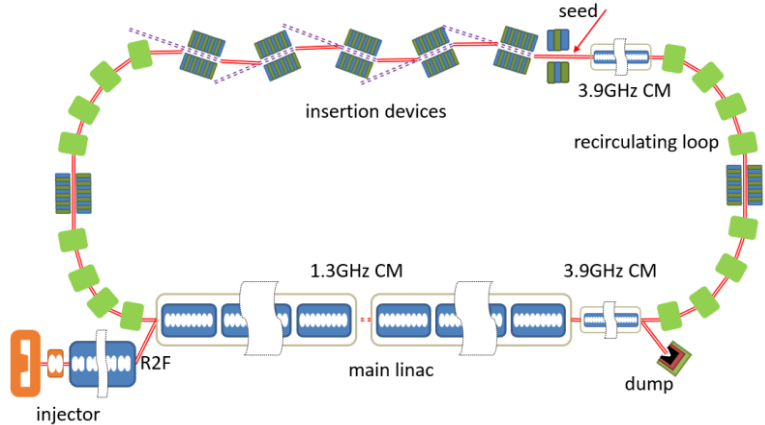

Fig. 1 the layout of the EUV light source. Round-to-flat (R2F) technique is adopted in the injector.

The injector consists of a 550 kV DC gun [26], a buncher, two solenoids and a cryomodule with 8 2-cell cavities. A 20 ps drive laser is used to generate the electron beam in the DC gun and then the electron beam is sent into the buncher to be compressed to about 4 ps (FWHM). After the acceleration in the cryomodule, a high-quality beam with the normalized emittance of 0.5 mm-mrad, energy of 15 MeV, peck current of 15 A and the pulse charge of 77 pC is achieved. By using the round-to-flat technique [27], the vertical emittance is reduced to 0.05 mm-mrad and the horizontal emittance is increased to about 5 mm-mrad at the exit of the injector. The main linac consists of six 1.3-GHz cryomodules and two 3.9-GHz cryomodules with eight 9-cell cavities in each module. The electron beam generated in the injector is boosted to about 600 MeV in the main linac and then transported through the recirculating loop. The recirculating loop comprises 4 TBA cells connected with one long straight section and two short straights, where the insertion devices are placed. Two operation modes, namely high-flux mode and high-resolution mode, can be provided via different setup of the linac and TBA cells. In the high-flux mode, a relative high peak current of the electron beam is generally required. To enhance the peak current, an additional energy chirp is achieved in the main linac. The nonlinear energy spread and the project energy chirp are compensated by the 3.9 GHz cavities in rear part of the linac and recirculating loop respectively. The electron beam is compressed to about 700 fs with the peak current of 100 A in the recirculating loop. In the high-resolution mode, a relatively long electron beam is needed. The electron beam is accelerated on-crest in the 1.3GHz cavities with nonlinear energy spread compensated by the 3.9 GHz cavities. The peak current maintains 15 A without compression in the recirculating loop. The beam parameters, as well as the machine parameters in the injector, linac and recirculating loop are listed in Table 1.

Table 1 beam parameters in the injector, linac and recirculating loop

| Parameters | Value | units |
|---|---|---|
| Beam energy (injector) | 15 | MeV |
| Beam energy (linac) | 600 | MeV |
| Normalized emittance (injector) | 0.5 | mm-mard |
| Normalized emittance (linac/undulator) | 5/0.05 | mm-mrad |
| Beam charge | 77 | pC |
| Pulse duration (linac, FWHM) | 4 | ps |
| Pulse duration (undulator, FWHM) | 0.7/4 | ps |
| Peak current | 100/15 | A |
| Relative energy spread | 0.1 | % |
| DC gun voltage | 550 | kV |
| Repetition rate | 1.3 | GHz |
| Drive laser duration | 20 | ps |
| Drive laser spot size (r) | 0.5 | mm |
| Bend angle in the ring | 30 | ° |

Three-dimensional (3D) numerical simulations were carried out to show the possible performances of the proposed EUV light source. The electron beam dynamics in the photo-injector was simulated with ASTRA [28], considering the space charge effects, under the help of genetic algorithm. The beam dynamics in the linac and the recirculating loop were carried out with ELEGANT [29] and MAD, considering the longitudinal wakefield effect and the coherent synchrotron radiation effects (CSR). The longitudinal phase space of the electron beam in the long drift line with different operation modes are shown in Fig. 2. The normalized emittance evolution along the half recirculating loop (2 TBA cells), the layout and optical function of one TBA cell are given in Fig. 3. The transverse emittances are well maintained as the beam passage through the injector, linac and recirculating loop parts, with the space charge, wakefield and CSR effects taking into account.

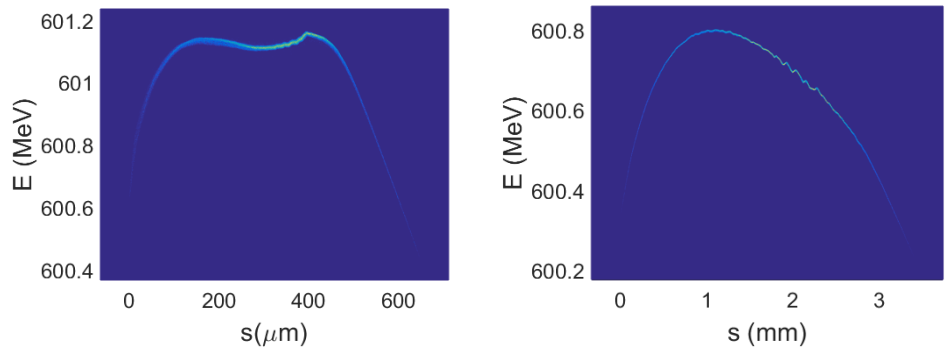

Fig. 2 longitudinal phase space of the electron beam in high-flux mode (left) and high-resolution mode (right) at the entrance of undulator

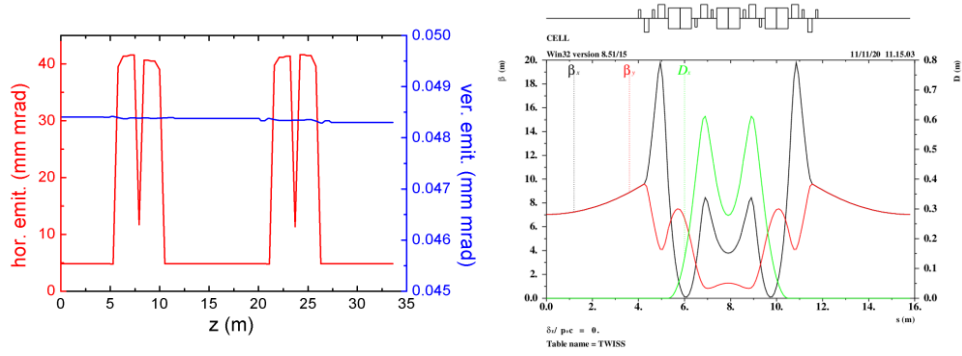

Fig. 3 emittance evolution (left) and the layout and optical function (right)

In the long straight section, the ADM scheme is adopted to generate coherent EUV radiation. After the electron beam passing through a bending magnet to get an angular dispersion, a seed laser with the wavelength of 257.5 nm, peak power of 10 kW and pulse length of 2 ps (FWHM) is sent into a 3-m-long modulator to interact with the electron beam to generate energy modulation. It is worth mentioning that the required laser power for ADM is about 3 orders of magnitude lower than that required for normal harmonic generation techniques [21, 22], making it possible to get the seed laser with the state-of-the-art techniques. The energy modulation is then converted into coherent microbunching as the beam passes through the downstream dogleg. The angles of all the bending magnets are set to be 0.2 rad and the distance between the bend centers of the dogleg is set to be 5 cm to optimize the bunching factor at 19$^{th}$ harmonic of the seed. One can find in Fig. 4b that the bunching factor at 19$^{th}$ harmonic is around 10%. Finally, fully coherent EUV radiation pulses at 13.5 nm can be generated in the downstream radiators. To support multiple-user operations, five 3-m-long radiators with period length of 2 cm are adopted in our design. With different angle of the undulator collimation and kick the electron beam after each radiator [30, 31], five different beamlines and end-stations can be supported. It is worth to point out here that the angle for a single kick is relatively small (25-50 μrad) based on the method used in Ref [30, 31]. To increase the offset angles of each radiator to mrad-level, especially designed bending systems are required [32, 33]. The modulator and radiator parts were simulated with GENESIS [34]. The simulation results with high-flux mode and high-resolution mode for one 3-m-long radiator are illustrated in Fig. 4 and Fig. 5. Based on the high-flux mode, coherent radiation pulses with peak power of 120 kW and spectral bandwidth of 3.5 meV can be achieved at the exit of each undulator. Considering the repetition rate of 1.3 GHz, the average output power is about 100 W (average brightness is calculated to be about $10^{25}$ phs/s/mm$^2$/mrad$^2$/0.1%BW). In the high-resolution mode, a 6 ps coherent radiation pulses can be generated with a spectral bandwidth of about 0.4 meV at the photon energy of 91.8 eV (13.5 nm). Higher resolution could be achieved by using longer electron beam and seed laser. This kind of GHz-level repetition rate coherent EUV light source with ~10$^9$ photons/pulse and sub-meV spectral resolution is highly required for experiments with angle-resolved photoemission spectroscopy (ARPES) techniques [35, 36]. Main parameters of the ADM and output coherent radiation are summarized in table2.

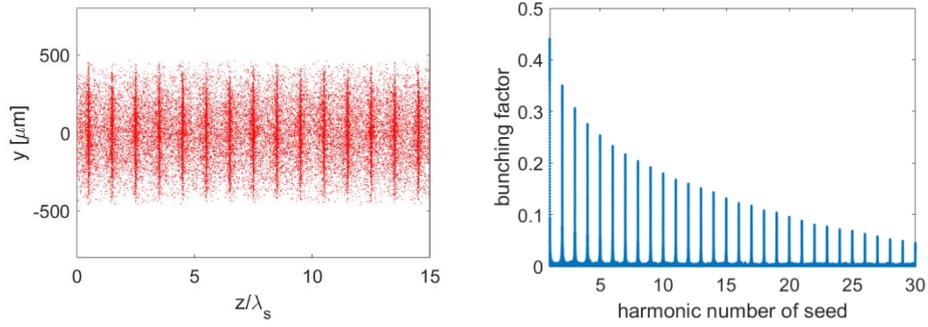

Fig. 4 simulation results of the density modulation and bunching factors at various harmonics of the seed laser.

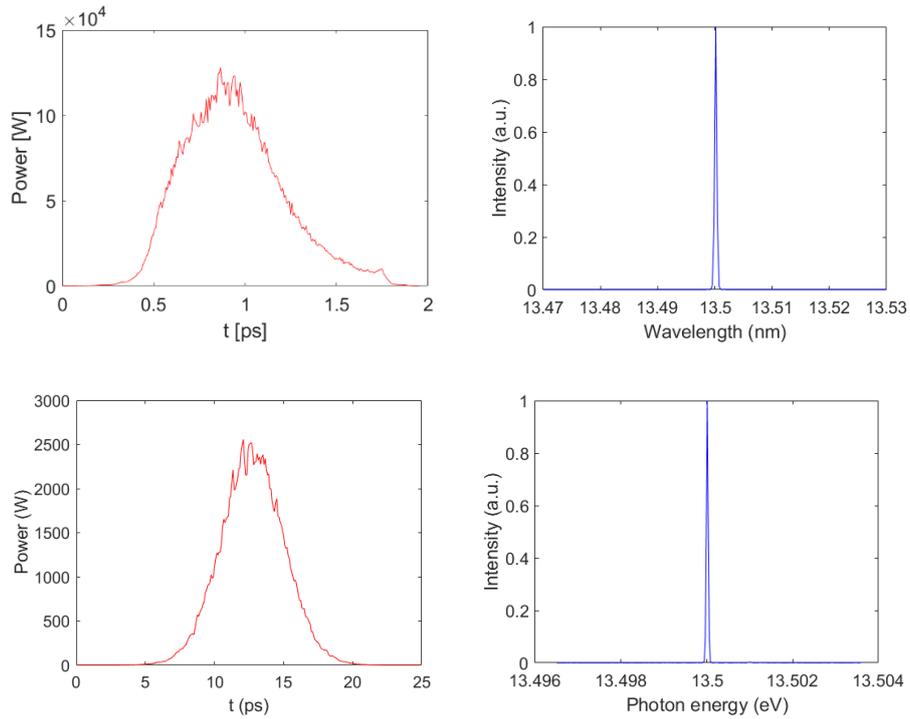

Fig. 5 radiation pulses and spectra at the exit of the radiator with high-flux mode (up) and high-resolution mode (down).

Table 2 simulation parameters for ADM

| Parameters | Value (high- flux/resolution mode) | units |
|---|---|---|
| Bending angle | 0.2 | rad |
| Modulator period | 3.5 | cm |
| Modulator length | 3 | m |
| Radiator period | 2 | cm |
| Radiator length | 3 | m |
| Seed laser wavelength | 257.5 | nm |
| Seed laser duration | 2/10 | ps |
| Seed laser peak power | 10 | kW |
| Radiation wavelength | 13.5 | nm |

| Radiation peak power | 120/2.5 | kW |
| Radiation pulse length | 0.7/6 | ps |
| Radiation pulse energy | 84/15 | nJ |
| Average output power | 100/19 | W |

After passing the long drift line, the electron beam is transported through the following half of the ring and then sent into the main linac to be decelerated to about 15 MeV. Finally, the electron beam is sent into the dump after the energy recovery process.

**Possible lattice design for ERL based full bandwidth light sources**

To further extend the wavelength coverage of the proposed technique, an ERL complex that consists of a low energy ERL and a high energy ERL is required. The low energy ERL, as we have discussed in the previous section, is used for generating coherent EUV seeding pulses for the high energy ERL. The injector for the high energy ERL is the same as the low energy one. High-quality electron beam with a bunch charge of 77 pC and emittance of 0.5 mm-rad is generated by the injector and then accelerated to about 3 GeV through an SRF main linac. The 3 GeV electron beam goes back to the entrance of the linac after being delivered by a recirculating loop and then passes through the linac again on deceleration phase for energy recovery. A schematic layout of the 3 GeV ERL light source is shown in Fig. 6 and the main beam parameters are listed in Table 3.

Table 3 main beam parameters of the 3 GeV single turn ERL light source

| Parameters | Value | units |
| --- | --- | --- |
| Beam energy | 3 | GeV |
| Normalized emittance | 0.5 | mm-mrad |
| Peak current | 15 | A |
| Beam charge | 77/8 | pC |
| Repetition rage | 1.3 | GHz |
| Average current | 100/10 | mA |

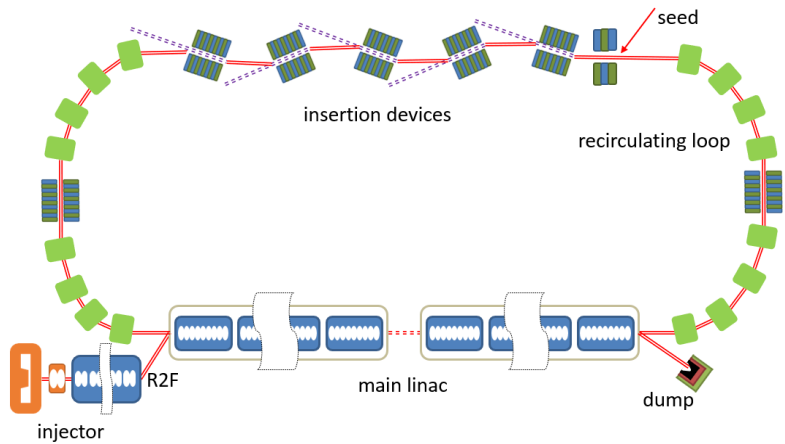

Fig. 6 schematic layout of a 3 GeV single turn ERL light source

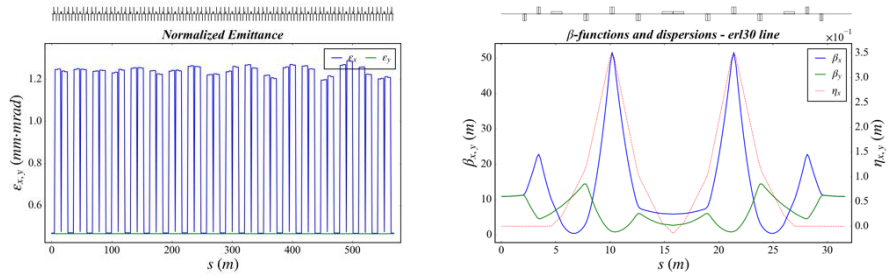

Fig. 7 the emittance evolution (left) and the optical function in each TBA cell (right)

The recirculating loop has a long straight section on the opposite side of the linac and two 180° arcs. Each arc comprises 18 periodical TBA cells which are isochronous. The optics functions of one TBA cell and the emittance evolution though the whole arc are shown in Fig. 7. The first arc transfers the electron beam from the end of the linac to the long straight section, which is kept for the coherent X-ray radiation generation.

By applying multi-turn acceleration and deceleration, the scale of the high energy ERLs can be reduced significantly. In our case, it is reasonable to use a double-turn ERL that integrates both the EUV and X-ray light sources. In such a scheme, the electron beam generated by the injector get the first 1.5 GeV energy gain on the first pass through the main linac. The EUV radiation is generated in a 1.5 GeV recirculating loop subsequently and then the electron beam passes through the linac again to have another 1.5 GeV energy gain. The recirculating loop for the 3.0 GeV electron beam is used for X-ray radiation generation. After that, the electron beam is decelerated twice in the linac for the energy recovery. It has to be point out that the main issue of this scheme is the maintenance of the beam quality as the electron beam passes through the 1.5 GeV ring. 3D simulations have been carried out and the results are given in Fig. 9. These results show that the quality of the electron beam can be well maintained, indicating the possibility of generating high repetition rate coherent x-ray pulses with the proposed technique.

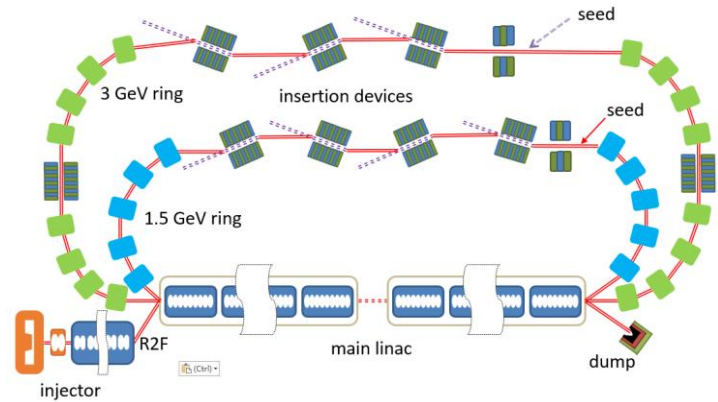

Fig.8 A possible layout for a fully coherent x-ray light source based on ERL.

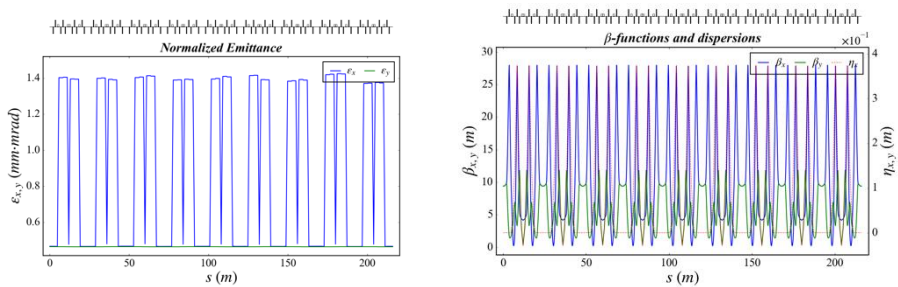

Fig. 9 the emittance evolution (left) and the optical function (right) along the 1.5 GeV ring.

**Conclusion**

In summary, a fully coherent light source that combines the ERL with ADM and round to flat beam techniques is proposed to generate high power EUV and x-ray radiation pulses. The average brightness of this light source is about 5-6 order of magnitude higher than that of the DLSRs with the same electron beam energy. Comparing with DLSRs and FELs, the proposed technique has the unique features of GHz-level high-repetition-rate and high-flux with a sub-meV spectral resolution. We believe that the proposed scheme will significantly enhance the capabilities of ERL based light sources and open up new opportunities for new science with ultrafast optics and spectroscopy techniques.